\documentclass[journal=jctcce,manuscript=article]{achemso}
\usepackage[utf8]{inputenc}
\usepackage[T1]{fontenc}
\usepackage[usenames,dvipsnames]{xcolor}
\usepackage[english]{babel}
\usepackage{graphicx}
\usepackage{rotating} 
\usepackage{float}
\usepackage{placeins}
\setkeys{acs}{doi = true}
\usepackage[colorlinks, linkcolor = blue, citecolor = blue, filecolor = blue,
urlcolor = blue]{hyperref}
\newcommand{\doi}[1]{\href{http://dx.doi.org/#1}{\nolinkurl{#1}}}
\usepackage{subcaption}
\usepackage{braket}
\usepackage{amsmath,amsfonts,mathtools}
\usepackage{caption}
\usepackage{cleveref}
\usepackage{comment}
\usepackage{cleveref}
\usepackage{color}

\title[Optimized MFML for QC]{Optimized Multifidelity Machine Learning for Quantum Chemistry}

\author{Vivin Vinod}
\affiliation{School of Mathematics and Natural Science, University of Wuppertal, 42119 Wuppertal, Germany}

\author{Ulrich Kleinekathöfer}
\affiliation{School of Science, Constructor University, Campus Ring 1, 28759 Bremen, Germany}

\author{Peter Zaspel}
\affiliation{School of Mathematics and Natural Science, University of Wuppertal, 42119 Wuppertal, Germany}
\email{zaspel@uni-wuppertal.de}

\begin{document}

\begin{abstract}
Machine learning (ML) provides access to fast and accurate quantum chemistry (QC) calculations for various properties of interest such as excitation  energies. It is often the case that high accuracy in prediction using an ML model, demands a large and costly training set. Various solutions and procedures have been presented to reduce this cost. These include methods such as $\Delta$-ML, hierarchical-ML, and multifidelity machine learning (MFML). 
MFML combines various $\Delta$-ML like sub-models for various fidelities according to a fixed scheme derived from the sparse grid combination technique. 
In this work we implement an optimization procedure to combine multifidelity models in a flexible scheme resulting in optimized MFML (o-MFML) that provides superior prediction capabilities. This hyper-parameter optimization is carried out on a holdout validation set of the property of interest.
This work benchmarks the o-MFML method in predicting the atomization energies on the QM7b dataset, and again in the prediction of excitation  energies for three molecules of growing size.
The results indicate that o-MFML is a strong methodological improvement over MFML and provides lower error of prediction. Even in cases of poor data distributions and lack of clear hierarchies among the fidelities, which were previously identified as issues for multifidelity methods, the o-MFML provides advantage to the prediction of quantum chemical properties.  
\end{abstract}

%

\section{Introduction}\label{Intro}
Fast and accurate calculations of chemical properties have become increasingly accessible to the community of quantum chemistry (QC) in the recent years with the accelerated development of machine learning (ML) for QC \cite{Huang2021, dral21a, westermayr_2021_perspective, dral2020quantum}. 
Various supervised and unsupervised learning approaches have seen widespread application in the field of QC. These applications include areas of material design and discovery \cite{Sergei21_Chem_review_NNML, westermayr_2021_perspective, Westermayr2020review, butler_davies_review_2018, lilienfeld2018_QML_in_CCS, Rupp_2018_dataenabled, ramprasad2017_MLinInMaterial, raccuglia2016ML_materialdiscovery, Pyzer2015_NeuralNetwork_materialDisc}
excitation energies \cite{vinod23_MFML, Cignoni2023, verm22a, dral21a, Westermayr2020NNKRR}, potential energy surfaces \cite{nandi2021delta, Chen21_CCSD_MLPotential, lin2020automatically, dral2020hierarchical, chmiela2018, chmiela2017, Natarajan15_PESwater}, and even prediction of chemical reactions \cite{Ahneman_chemicalReactions_ML} and ML molecular dynamics for the simulation of infrared spectra \cite{Infrared_spectra_gastegger}.
The conventionally costly QC calculations are gradually being replaced with ML models or hybrids of ML and QC resulting in a drastic reduction of the compute cost associated with chemical design and discovery. The core principle of the various ML techniques is to reproduce some implicit mapping between the geometry of the molecules to some property of interest such as excitation  energies, potential energy surfaces, or atomization energies.
These are usually targeted at some level of theory which is relevant to the area of application. 

The general ML-QC pipeline for such applications begins with the generation of raw data consisting of the Cartesian geometries of the molecules of interest and the QC calculation property to be predicted at the level of theory (MP2, CCSD etc) that is deemed accurate for the application. 
The Cartesian coordinates are then transformed into some input feature format, called \textit{representations} or \textit{molecular descriptors}, that the ML models can map to the property of interest. In the recent past, much work has been dedicated to the development of such representations. These include molecule-wise descriptors such inverse distance representations and their extensions such as the Coulomb Matrix (CM) \cite{schutt2018schnet, rama15a, rupp2014machine,  montavon2013machine, Rup12CM,  montavon2013machine, hansen2013assessment} and Bag of Bonds \cite{de2016comparing, hansen2015_BoB, bartok2013representingSOAP}, or atom-wise descriptors such as Smooth Overlap of Atomic Positions (SOAP) \cite{bartok2013representingSOAP, bartok2010gaussian}, SLATM \cite{Huang2020slatm}, permutationally invariant polynomials (PIP) \cite{nandi2021delta}, the PaiNN representation \cite{schutt21a_PAINN}, and the Faber-Christensen-Huang-Lilienfeld (FCHL) representation \cite{Huang2021, Christensen2020, montavon2013machine, Rup12CM}. Significant research has also been performed on using other types of representations such as SMILES strings \cite{Kang20_SMILES_application, Weininger88_SMILES}, graph-based representations \cite{david2020moleculardescriptors}, and representations that are either generated with neural network (NN) models such as the Deep Tensor NN \cite{Schutt19_NNrepresentation, schutt2018schnet, schutt2017quantum} or are generated \textit{ad hoc} \cite{carrete2014finding, pilania2013accelerating}. Once machine interpretable features are generated, any of the various ML methods such as kernel ridge regression (KRR), Gaussian Process Regression (GPR), or NN models such as ANI \cite{Gao20_TorchAni, smit17_ANI-1}, SchNet \cite{schutt2018schnet, schutt2017quantum} and PhysNet \cite{unke2019physnet}, can be used to map the input features to their respective QC properties. 

Within such frameworks, it has been a common observation that the higher the number of training samples, the better the accuracy of the prediction.
However, a high cost is associated with generating this training data since conventional QC calculations with high accuracy are expensive to generate.
Thus, the compute cost associated with discovery in QC is shifted from the conventional QC calculations to the cost associated with generating the training data set for these ML models.
While any of the aforementioned ML methods is a promising candidate to replacing the time consuming conventional calculations, only rather recently has the cost of the training data for the models been investigated \cite{vinod23_MFML, dral2020hierarchical, zasp19a, Sun2019_deltaML_NN}. 
Various techniques and models have been previously implemented to reduce this overhead cost.
Among these are methods such as the $\Delta$-ML \cite{NN_deltaml_Liu22, Ruth22_CCSD_deltaML, nandi2021delta, patra2020multi, Pilania2017, Ramakrishnan2015}, and active learning approaches \cite{bernstein2019novo, behler2015constructing}. An \textit{ad hoc} optimization procedure for the $\Delta$-ML method has been implemented for the ground state PES reconstruction of the $\rm CH_3Cl$, termed as hierarchical-ML (h-ML) \cite{dral2020hierarchical}. Based on the CPU compute time of point calculations, the training samples to be used at various fidelities are selected by minimizing an objective function. This reduces the number of QC calculations needed to generate the multifidelity data set for some user defined target error.

Recently, a systematic generalization of the $\Delta$-ML method called multifidelity machine learning (MFML) \cite{zasp19a} was applied to the first excitation  energies of molecules \cite{vinod23_MFML}. 
The MFML method exploits the existence of varying levels of accuracy of conventional QC methods, thereby resulting in a hierarchy of methods for properties such as excitation energies.
MFML reduces the number of expensive training samples needed by training on the difference of various \textit{fidelities} between a \textit{baseline fidelity} and the \textit{target fidelity}. The MFML model is built by iteratively adding models built on the difference between the excitation  energies calculated at the various fidelities. 
In MFML the number of training samples is decreased by 2 at each subsequently costly fidelity \cite{vinod23_MFML}. Thus, there is an inherent decrease in the number of costly training samples.
For each fidelity and training set size at this corresponding fidelity, a \textit{sub-model}, for a given training set size, is trained \cite{zasp19a}. This is recursively performed from a \textit{baseline} fidelity (cheaper and less accurate) up to the target fidelity (expensive and more accurate).
The various sub-models are combined to give the final MFML model. 
This combination was performed based on the sparse-grid combination technique \cite{hegland2016combination, haji2016multi, haji2016multi, harbrecht2013combination, reisinger2013combination, benk2012hybrid, Reisinger2012} as has been discussed in Refs.~\cite{vinod23_MFML, zasp19a}.  

This work furthers the methodological research in MFML by introducing a novel method of optimally combining the various sub-models built on the different fidelities.
The novel approach is inspired by Refs.~\cite{hegland2007combination, garcke2006regression} where an optimized sparse-grid combination technique is introduced and discussed for the solution of partial differential equations. In contrast to that work, we however apply it to ML for QC where
the optimal combination of the sub-models is performed with respect to a validation set of the property of interest, not based on intrinsic approximation properties of the given problem. This results in a multifidelity model that predicts the property at the target fidelity with improved accuracy (Section \ref{Results}).
Thus, the optimized MFML (o-MFML) presents an optimal linear combination of the sub-models. This work benchmarks this novel method on the QM7b dataset with the prediction of atomization energies at the CCSD level of theory with the ccpvdz basis set \cite{zasp19a, montavon2013machine}. Further benchmarking is carried out on the first excitation  energy data-set from Ref.~\cite{vinod23_MFML}.
The results indicate that the o-MFML is indeed superior to the implementation of the conventional MFML. 

The manuscript is structured as follows. A brief overview of the data used for this study is reported in Section \ref{Dataset}. Next, Section \ref{Methods} discusses the key methodology and the novel o-MFML technique. Next, various results of the comparison of MFML and o-MFML for the two datasets are delineated. Section \ref{qm7bresults} discusses the results for the benchmark on the QM7b dataset \cite{montavon2013machine, montavon2013machine} while Section \ref{excitedstate_results} discusses the corresponding results for the excitation  energy predictions. 
The assessment of the various models is carried out by studying the mean absolute error (MAE) and the learning curves.

\section{Methods}\label{Methods}
In this section, the various methodological terms needed to arrive at the results are recorded. Details of dataset, MFML definitions, and optimization methods are discussed in addition to the evaluation metric for the various ML models.

\subsection{Dataset}\label{Dataset}
The effectiveness of the optimized MFML method is benchmarked on the QM7b dataset \cite{montavon2013machine}, which consists of a total of 7211 molecules with up to seven heavy atoms. 
The atomization energies for each of these molecules were calculated in kcal/mol as mentioned in Ref.~\cite{zasp19a}. For this study, the effective averaged atomization energies are considered. This is given as:
\begin{equation}
    E_{\rm eff} := E - \sum_{i}n_i \cdot e_i~,
    \label{eq_eff_at_energy}
\end{equation}
where, $n_i$ is the number of atoms and $e_i$ is the effective atomic energy of the $i^{\rm th}$ molecule. The latter is obtained by a linear fit of $E=\sum_i n_i \cdot e_i$ for all molecules in the QM7b dataset. 
Without loss of generality, the $E_{\rm eff}$ used are simply referred to as atomization energies herein.
Further, only the MP2 \cite{Yost_MP2_theory, Quin_MP2_DFT_theory, Pogrebetsky_MP2_theory} and CCSD \cite{Purvis_CCSD_theory, Bartlett_CCSD_theory, Crawford_CCSD_theory} levels of theory were considered. 
The fidelity structure was formed by evaluating these with three varying basis set sizes, namely: STO-3G, 6-31G, and ccpvdz (with increasing size). 
While the original use of this dataset in Ref.~\cite{zasp19a} considers a 3-dimensional multifidelity structure, in this work these are flattened into a 2-dimensional multifidelity structure. 
Thus the order of the fidelities in the assumed hierarchy was taken as MP2-STO3G, MP2-631G, MP2-ccpvdz, CCSD-STO3G, CCSD-631G, and CCSD-ccpvdz. The CCSD-ccpvdz is set as the target fidelity. A total of $1.5\cdot2^7=6144$ molecules were randomly chosen as the training set.

The data for the excitation energy calculations is taken from Ref.~\cite{vinod23_MFML}. This consists of DFTB and MD based simulation of benzene, naphthalene, and anthracene. For each, a total of 15~ps of trajectory was generated after energy minimization and equilibration. This trajectory was then sampled every 1~fs giving 15,000 frames which was used for training and evaluation. For training, the first $N_{\rm train}=1.5\cdot2^{13}=12288$ frames were used with excitation  energies calculated at five fidelities (basis sets): def2-TZVP, def2-SVP, 6-31G, 3-21G, and STO-3G. The sampling and calculations are identical to those discussed in Ref.~\cite{vinod23_MFML}.

\subsection{Multifidelity Machine Learning} \label{MFML_methods}
Consider an ordered hierarchy of fidelities indexed as $f=1,2,\ldots, F$ where the cost of calculation (and usually, therefore, accuracy) increases with an increase in the index. The training set for data at some fidelity $f$ can be then defined as $\mathcal{T}^{(f)}:=\left\{\left(\boldsymbol{X}^{(f)}_i,y_i^{(f)}\right) \right\}_{i=1}^{N^{(f)}}$.
Defining the set of molecular descriptors $\mathcal{X}^f := \left\{\boldsymbol{X}_i^f\lvert \left(\boldsymbol{X}_i^f,E^f_i\right)\in\mathcal{T}^{(f)}\right\}$, based on previous work in this field as detailed in Refs.~\cite{vinod23_MFML, zasp19a} the current state of the multifidelity method recommends the nestedness $\mathcal{X}^F\subseteq \ldots \subseteq \mathcal{X}^2 \subseteq \mathcal{X}^1$ of the training data. This is enforced in both the datasets used in this work.
That is, if a molecular conformation is picked, which has the quantum chemistry property calculated at the highest fidelity, then it is also that the quantum chemistry property is calculated for this conformation at the next lower fidelity, and so on.
As Ref.~\cite{vinod23_MFML} shows, a multifidelity machine learning (MFML) model with kernel ridge regression (KRR) as the ML model of choice, can be iteratively built for an ordered hierarchy of fidelities as
\begin{equation}
    P^{(F;f_b)}_{\rm MFML}\left(\boldsymbol{X}_q\right) := P^{(f_b)}_{\rm KRR}\left(\boldsymbol{X}_q\right) + \sum_{f_b\leq f<F}P_{\rm KRR}^{(f,f+1)}\left(\boldsymbol{X}_q\right)~,
    \label{eq_MFML_og}
\end{equation}
where $F$ is the target fidelity and $f_b=1,2,\ldots ,F-1$ is some baseline fidelity, and $\boldsymbol{X}_q$ is the representation of a query molecule.The term inside the summation is calculated as
\begin{equation}
    P_{\rm KRR}^{(f,f+1)}\left(\boldsymbol{X}_q\right) := \sum_{i=1}^{N_{\rm train}^{(f+1)}}\alpha_i^{(f,f+1)} k\left(\boldsymbol{X}_i,\boldsymbol{X}_q\right)
    \label{eq_MFML_singleKRR}~. 
\end{equation}
The coefficients of KRR, $\alpha_i^{(f,f+1)}$, are calculated by solving the linear system of equations given by
\begin{equation}
    \left(\boldsymbol{K}+\lambda \boldsymbol{I}_{N^{(f+1)}}\right)\boldsymbol{\alpha}^{(f,f+1)} = \boldsymbol{\Delta y}^{(f,f+1)}
    \label{eq_MFML_alphas}~.
\end{equation}
It is to be noted that $\boldsymbol{\Delta y}^{(f,f+1)}=\boldsymbol{y}^{f+1}-\boldsymbol{y}^{(f,f+1)}$,
where $\boldsymbol{y}^{f+1}$ is the vector of energies in the training set $\mathcal{T}^{(f+1)}$ and $\boldsymbol{y}^{(f,f+1)}$ is the vector of energies in training set $\mathcal{T}^{(f)}$ restricted to those conformations only found on fidelity level $f+1$. 
Thus, this definition of MFML can be seen as one that works on the difference between the data. 
As an example, for a target fidelity $F=5$, with a baseline fidelity $f_b=1$, the MFML model built with Eq.~\eqref{eq_MFML_og} would be explicitly written as:
$$P_{\rm MFML}^{(5;1)} = P_{\rm KRR}^{(1)} + P_{\rm KRR}^{(1,2)} + P_{\rm KRR}^{(2,3)} + P_{\rm KRR}^{(3,4)} + P_{\rm KRR}^{(4,5)}~.$$
The number of training samples used for each of the fidelities is scaled by 2, based on work in MFML \cite{vinod23_MFML, zasp19a}. Thus, if the number of training samples at the target fidelity are set to be $N_{\rm train}^{F}$, then the next lower fidelity uses $2\cdot N_{\rm train}^{F}$ of training samples and so on.

Ref.~\cite{zasp19a} has mathematically shown that this form of the MFML is equivalent to taking the difference of models built on the two different levels while ensuring a nested data structure. 
That is, $P_{\rm KRR}^{(f,f+1)}\equiv P_{\rm KRR}^{(f+1)} - P_{\rm KRR}^{(f)}$ where $P_{\rm KRR}^{(f)}$ is built on $\left\{\boldsymbol{X}_i,y_i^{(f,f+1)}\right\}_{i=1}^{N_{\rm train}^{(f+1)}}$ with conformations restricted to those found in the training set used for fidelity $f+1$. 
This is further numerically verified in the supplementary material in S~3.1 for the first excitation energy data.
Models of the type $P_{\rm KRR}^{(f+1)}$ and $P_{\rm KRR}^{(f)}$ are herein referred to as \textit{sub-models} of MFML. A sub-model of MFML is built for a specific choice of a training set. For the current work, it implies selecting a fidelity, $f$, and the number of training samples at this fidelity, $N_{\rm train}^{(f)}$ for $f=1,\ldots,F$. This formulation of sub-models, represents a 2-dimensional multifidelity structure, that is, the fidelity, and the number of training samples. 
In such a structure, it is assumed that increasing the fidelity results in a more accurate (and therefore, a costlier) QC calculation. This in turn translates into a more accurate (and costlier to train) sub-model. 
In principle, there is no limit on the dimensions of MFML as long as a clear hierarchy can be established in each dimension \cite{zasp19a}. 
For the specific case of the 2-D structure, one can identify a sub-model with an ordered pair, or index, $\boldsymbol{s}=(f,\eta_f)$ where $f$ is the fidelity and the number of training samples chosen from this fidelity are given as $N_{\rm train}^{f} = 2^{\eta_f}$. 
A standard KRR model (see Section S1) built for the index $\boldsymbol{s}$ is then denoted as $P_{\rm KRR}^{(\boldsymbol{s})}$. 
The conceptual development of such a combination of sub-models has been previously implemented by Zaspel \textit{et al.} for the prediction of atomization energies in the QM7b dataset \cite{zasp19a}. 

With this development, one arrives at the MFML method written as the linear combination of the various sub-models. To this end, some notations are introduced. The set of indexes of all available sub-models is denoted by $\mathcal{S}$. A standard KRR model for a query molecule represented as $\boldsymbol{X}_q$ is built as $P_{\rm KRR}^{(\boldsymbol{s})}\left(\boldsymbol{X}_q\right)$ for $\boldsymbol{s}\in \mathcal{S}$. Further, define the set of indexes of sub-models used for a MFML model with target fidelity $F$, for $N_{\rm train}^{F} = 2^{\eta_F}$, and a baseline $f_b$, as follows:
\begin{multline}    
    \mathcal{S}^{\left(F,\eta_F;f_b\right)}:=  \Big{\{}(f,\eta_f)\in\mathcal{S} \lvert
    f \in\left\{f_b,\ldots,F\right\},
     \eta_f\in\{\eta_F,\ldots,2^{F-f_b}\cdot \eta_F \},\\
    F+\eta_F-1\leq f+\eta_f \leq F+\eta_F \Big{\}}
    \label{eq_subspace_MFML}~,
\end{multline}
where $\mathcal{S}^{(F,\eta_F;f_b)}\subseteq \mathcal{S}$. 
The motivation is to combine various sub-models such that only a few expensive training samples are required, which, when combined with cheaper training samples, yield a high-accuracy low-cost model for the target fidelity. This is achieved by the linear combination of the sub-models from $\boldsymbol{s}\in S^{(F,\eta_F;f_b)}$. 
This is denoted by
\begin{equation}
    P_{\rm MFML}^{(F,\eta_F;f_b)}\left(\boldsymbol{X}_q\right) := \sum_{\boldsymbol{s}\in\mathcal{S}^{(F,\eta_F;f_b)}} \beta_{\boldsymbol{s}} P^{(\boldsymbol{s})}_{\rm KRR}\left(\boldsymbol{X}_q\right)~,
    \label{eq_MFML_linearsum}
\end{equation}
where  
$\beta_{\boldsymbol{s}}$ are the coefficients of the linear combination. These coefficients can be interpreted as a measure of how much each sub-model contributes to the final MFML model. 
Based on work in MFML for atomization energies \cite{zasp19a} and excitation  energies \cite{vinod23_MFML}, the coefficients are set in such a manner that each sub-model contributes in equal magnitude to the final MFML model. For a model of the form $P_{\rm MFML}^{(F,\eta_F;f_b)}$, the $\beta_{\boldsymbol{s}}$, are set in conventional MFML as follows:
\begin{equation}
    \beta_{\boldsymbol{s}}^{\rm MFML} = \begin{cases}
        +1, & \text{if } f+\eta_f = F+\eta_F\\
        -1, & \text{otherwise}
    \end{cases}~,
    \label{eq_MFML_beta_i}
\end{equation}
where the terms are as discussed previously.
\begin{figure}[htb]
    \centering
    \includegraphics[width=0.6\textwidth]{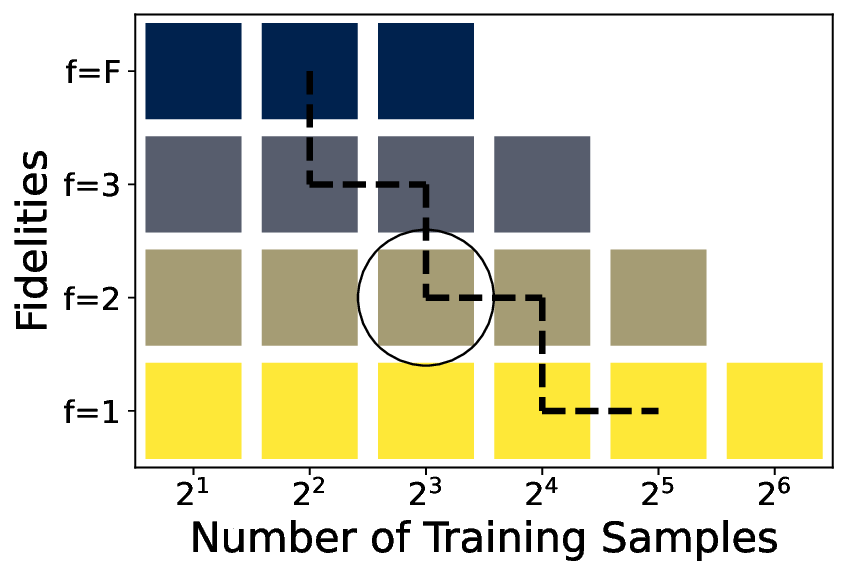}
    \caption{A hypothetical structure of sub-models for 4 fidelities is depicted here. Each sub-model can be identified with an index pair $\boldsymbol{s}=(f,\eta_f)$ representing the fidelity with $N_{\rm train}^{f}=2^{\eta_f}$. Thus the circled sub-model can be denoted as $\boldsymbol{s'}=(2,3)$. Within this formulation, the MFML model is built by combining the sub-models as shown with the dotted black line. The contribution of sub-model $\boldsymbol{s'}$ is given by the coefficient denoted by $\beta_{\boldsymbol{s'}}$. In conventional MFML, this would in particular be -1.
    }
    \label{fig_subspaces}
\end{figure}

A hypothetical 2-dimensional multifidelity structure is shown in Figure \ref{fig_subspaces} with the dimension of fidelity on the y-axis and the dimension of the number of samples on the x-axis.
One can now identify various sub-models in this hypothetical structure. For example, $P^{(\boldsymbol{s})}$ with $\boldsymbol{s}=(6-31G, 5)$, represents a sub-model built at the 6-31G fidelity with $2^5=32$ training samples. 
In this scheme, the cost (and therefore, the accuracy to target fidelity) of the training data of the sub-models increases with increase in either of $f$ or $s\eta_f$. 
That is, $\boldsymbol{s}$ is more accurate (and more expensive) than a sub-model built with with $\boldsymbol{s'}=(3-21G,5)$. At the same time, a sub-model built with $\boldsymbol{s''}=(6-31G, 6)$ is more accurate (and expensive) than $\boldsymbol{s}$ from this example. 
As an example to depict the conventional MFML model built with the various sub-models, consider the set of sub-models for MFML being built for target fidelity $F=4$, with $2^2$ (that is, $\eta_F = 2$) training samples at this fidelity, and with a baseline fidelity of $f_b=1$. 
The set of MFML sub-model indexes is then given as $\mathcal{S}^{\left(4,2;1\right)} = \left\{ (4,2),(3,2),(3,3),(2,3),(2,4),(1,4),(1,5)\right\}$. The MFML model is built as the linear combination of the individual KRR sub-models with indexes $s\in\mathcal{S}^{\left(4,2;1\right)}$. The coefficients are as defined by Eq.~\eqref{eq_MFML_beta_i}. Explicitly, 
\begin{multline*}
    P_{\rm MFML}^{\left(4,2;1\right)}\left(\boldsymbol{X}_q\right) 
    := P^{((4,2))}_{\rm KRR}\left(\boldsymbol{X}_q\right)-P^{((3,2))}_{\rm KRR}\left(\boldsymbol{X}_q\right) 
    + P^{((3,3))}_{\rm KRR}\left(\boldsymbol{X}_q\right)-P^{((2,3))}_{\rm KRR}\left(\boldsymbol{X}_q\right)
    +\\ P^{((2,4))}_{\rm KRR}\left(\boldsymbol{X}_q\right)-P^{((1,4))}_{\rm KRR}\left(\boldsymbol{X}_q\right) 
    + P^{((1,5))}_{\rm KRR}\left(\boldsymbol{X}_q\right)
    ~.
\end{multline*} 
One can readily see that this is the very same model $P_{\rm MFML}^{(4;1)}\left(\boldsymbol{X}_q\right)$ as would be arrived at by using Eq.~\eqref{eq_MFML_og} with $2^2$ training samples used at the target fidelity. 
The conventional MFML model built with coefficients set by Eq.~\eqref{eq_MFML_beta_i} is simply denoted as $P_{\rm MFML}^{(F;f_b)}$ since it is identical to the MFML model built in Ref.~\cite{vinod23_MFML}.

\subsection{Optimized MFML} \label{POM_details}
Having written the MFML model in terms of the individual sub-models of multifidelity, one can consider formulations of the coefficients, which are different from Eq.~\eqref{eq_MFML_beta_i}. 
This can be seen as a hyper-parameter optimization of the different $\beta_{\boldsymbol{s}}$ to return a multifidelity model which has improved accuracy at the target fidelity.
In interest of this form of an optimization, the validation set is defined as 
$\mathcal{V}^F_{\rm val}:=\{(\boldsymbol{X}_q^{\rm ref},E^{\rm ref}_q)\}_{q=1}^{N_{\rm val}}$. 
To evaluate the accuracy of the model, define a test set $\mathcal{V}^F_{\rm test}:=\{(\boldsymbol{X}_q^{\rm ref},E^{\rm ref}_q)\}_{q=1}^{N_{\rm test}}$ such that 
the two are mutually exclusive. That is,
$\mathcal{V}^F_{\rm val} \cap \mathcal{V}^F_{\rm test} = \phi$, where $\phi$ denotes the empty set. 
The split of the validation and test sets is a common approach in ML techniques wherein the optimization/ hyperparameter-tuning is performed on the former and the error of the final model is reported on the latter. It is to be noted that the test set is never used in any stage of the training process. 

One can explicitly define an optimized MFML (o-MFML) model for a target fidelity $F$, with $N^{(F)}=2^{\eta_F}$ training samples at the target fidelity, for a baseline fidelity $f_b$, as 
\begin{equation}
    P_{\rm o-MFML}^{\left(F,\eta_F;f_b\right)}\left(\boldsymbol{X}_q\right) := 
    \sum_{\boldsymbol{s}\in \mathcal{S}'}\beta_{\boldsymbol{s}}^{\rm opt} P^{(\boldsymbol{s})}_{\rm KRR} \left(\boldsymbol{X}_q\right)
    \label{eq_POM_def}
\end{equation}
where $\beta_{\boldsymbol{s}}^{\rm opt}$ are optimized coefficients, and $\boldsymbol{X}_q$ is the representation of a query molecule. 
In general, one is interested in solving the optimization task:
$$
    \beta_{\boldsymbol{s}}^{\rm opt} = \arg\min_{\beta_{\boldsymbol{s}}} 
    \left\lVert \sum_{v=1}^{N_{\rm val}} \left(y_v^{\rm ref} - \sum_{\boldsymbol{s}\in S'} \beta_{\boldsymbol{s}} P^{(\boldsymbol{s})}_{\rm KRR}\left(\boldsymbol{X}_v\right)\right) \right\rVert_p 
$$
where one minimizes some $p$-norm on the validation set $\mathcal{V}^F_{\rm val}$. This is equivalent to solving 
\begin{equation}
    \boldsymbol{\beta}^{\rm opt} = \arg\min_{\boldsymbol{\beta}} 
    \left\lVert
    \boldsymbol{M}_{\mathcal{S'}}\boldsymbol{\beta} - \boldsymbol{y}^{\rm ref}
    \right\rVert_p
    \label{eq_opt_min_problem}
\end{equation}
where $\boldsymbol{M}_{\mathcal{S}'} = \left(P_{\rm KRR}^{(\boldsymbol{j})}\left(\boldsymbol{X}_i\right)\right)_{i=1,\ldots,N_{\rm val};\boldsymbol{j}\in\mathcal{S}'}$ is a $N_{\rm val}\times \lvert\mathcal{S}'\rvert$ matrix, $\boldsymbol{\beta}$ is the vector of coefficients with respect to $\mathcal{S}'$ as depicted in Eq.~\eqref{eq_POM_def}, and $\boldsymbol{y}^{\rm ref}$ is the vector of reference energies from $\mathcal{V}^F_{\rm val}$.
This work utilizes the ordinary least square optimization (OLS) procedure to solve Eq.~\eqref{eq_opt_min_problem} with $p=2$.
In the results, the OLS optimized MFML model is reported as $P_{\rm o-MFML}$. However, it must be noted that any method that can solve the minimization problem in Eq.~\eqref{eq_opt_min_problem} can be used to optimize the coefficients. 

Thus, the complete process of building an o-MFML model can be written as follows:
\begin{enumerate}
    \item Identify the set of sub-models for a given MFML model, $\mathcal{S}^{(F,\eta_F;f_b)}$.
    \item Build the various KRR sub-models for sub-models $\boldsymbol{s}\in \mathcal{S}^{\left(F,\eta_F;f_b\right)}$.
    \item Optimize the coefficients, $\beta_{\boldsymbol{s}}$, on $\mathcal{V}^F_{\rm val}$ using an optimizer of choice.
    \item Evaluate the final model $P_{\rm o-MFML}^{(F,\eta_F;f_b)}$ on $\mathcal{V}^F_{\rm test}$ for some error metric (Section \ref{ModelEval}).
\end{enumerate}

\subsection{Model Evaluation} \label{ModelEval}
Learning curves are a well known metric in the field of KRR-based ML methods. These depict the change in prediction error of the model for increasing training set size.
In all results reported in this work, the learning curves are averaged over a 10-run random shuffling of the MFML training set while ensuring the nestedness of the training samples. 
For each of the 10 runs, the procedure is as follows:
\begin{enumerate}
    \item Randomly select $N_{\rm train}^{F}=2^{\eta_F}$ training samples from $\mathcal{T}^F$. Define this as a new sampled training set, $\mathcal{G}^{F}\subseteq\mathcal{T}^{F}$.
    \item Train the sub-model $P_{\rm KRR}^{(F,\eta_{F})}$ on training data from $\mathcal{G}^{F}$.
    \item For the conformations $\boldsymbol{X}_i$ such that $\left(\boldsymbol{X}^F_i,y_i^{F}\right)\in\mathcal{G}^F$, train the sub-model $P_{\rm KRR}^{(F-1,\eta_{F})}$ with properties $y_i^{F-1,F}$, that is, the energies at fidelity $F-1$ for the conformations which are also found in $\mathcal{G}^{(F)}$. 
    \item At the next lower fidelity, $f=F-1$,
    build the sampled training set $$\mathcal{G}^{F-1}:= \left\{\left(\boldsymbol{X}_i^{F},y_i^{F-1}\right)\right\}_{i=1}^{N_{\rm train}^{F}}\cup\left\{\left(\boldsymbol{X}_j^{F-1},y_j^{F-1}\right)\right\}_{j=1}^{2\cdot N_{\rm train}^{F} - N_{\rm train}^{F}}~,$$
    such that $\left\{\left(\boldsymbol{X}_i^{F},y_i^F\right)\right\}\in\mathcal{G}^F$ and $\left\{\left(\boldsymbol{X}_j^{F-1},y_j^{F-1}\right)\right\}\in\mathcal{T}^{F-1}\setminus\mathcal{G}^{F}$ is randomly sampled.
    \item Train the sub-model $P_{\rm KRR}^{F-1,\eta_{F-1}}$ on $\mathcal{G}^{F-1}$. Similar to step 3, train $P_{\rm KRR}^{(F-2,\eta_{F-1})}$.
    \item Repeat steps 4 and 5 recursively until baseline fidelity, $f=f_b$.
\end{enumerate}
Throughout this investigation, all prediction errors have been reported on a test set, $\mathcal{V}_{\rm test}^F:=\{(\boldsymbol{X}_q^{\text{ref}},y^{\text{ref}}_q)\}_{q=1}^{N_{\rm test}}$, which consist of evaluation representations and their corresponding reference values for property of interest (for example, excitation  energy) calculated at the target fidelity $F$ (for example, TZVP). 
These errors are reported as Mean Absolute Errors (MAEs) which are defined by a discrete $L_1$ norm
\begin{equation}
    MAE = \frac{1}{N_{\rm test}}\sum_{q=1}^{N_{\rm test}}\left\lvert P_{\rm ML}\left(\boldsymbol{X}_q^{\text{ref}}\right) - {y}^{\text{ref}}_q\right\rvert~.
    \label{eq_MAE}
\end{equation}
The model $P_{\rm ML}$ can be either identified by the standard KRR model or by the various MFML models discussed in this work. 
For the case of predicting atomization energies for the QM7b dataset, of the 1067 molecules which remained after separating the training data, 367 were randomly sampled and used as the validation set along with their atomization energies calculated at the CCSD-ccpvdz fidelity. The remaining 700 molecules and their atomization energies at the target fidelity were utilized as the test set.
In the case of the excitation  energy dataset, for each molecule, the 2712 samples with the target fidelity of TZVP were randomly split into 712 and 2000 samples for the validation and test set respectively. The random sampling was performed using the Scikit-learn package \cite{scikit-learn}. 

\section{Results}\label{Results}
To establish the effectiveness of the optimized MFML (o-MFML) method, a study was carried out on two datasets for the prediction of two different properties. 
In particular, this work reports the prediction of atomization energies for the QM7b dataset as calculated in Ref.~\cite{zasp19a}, and the prediction of the first excitation  energies for the data used in Ref.~\cite{vinod23_MFML}. 
The process of the kernel generation and training of the KRR for the work recorded here are carried out with the QML package \cite{Christensenqmlcode}. 

\subsection{Atomization Energy Prediction on QM7b}\label{qm7bresults}
\begin{figure}[htb!]
    \centering
    \includegraphics[width=0.75\textwidth]{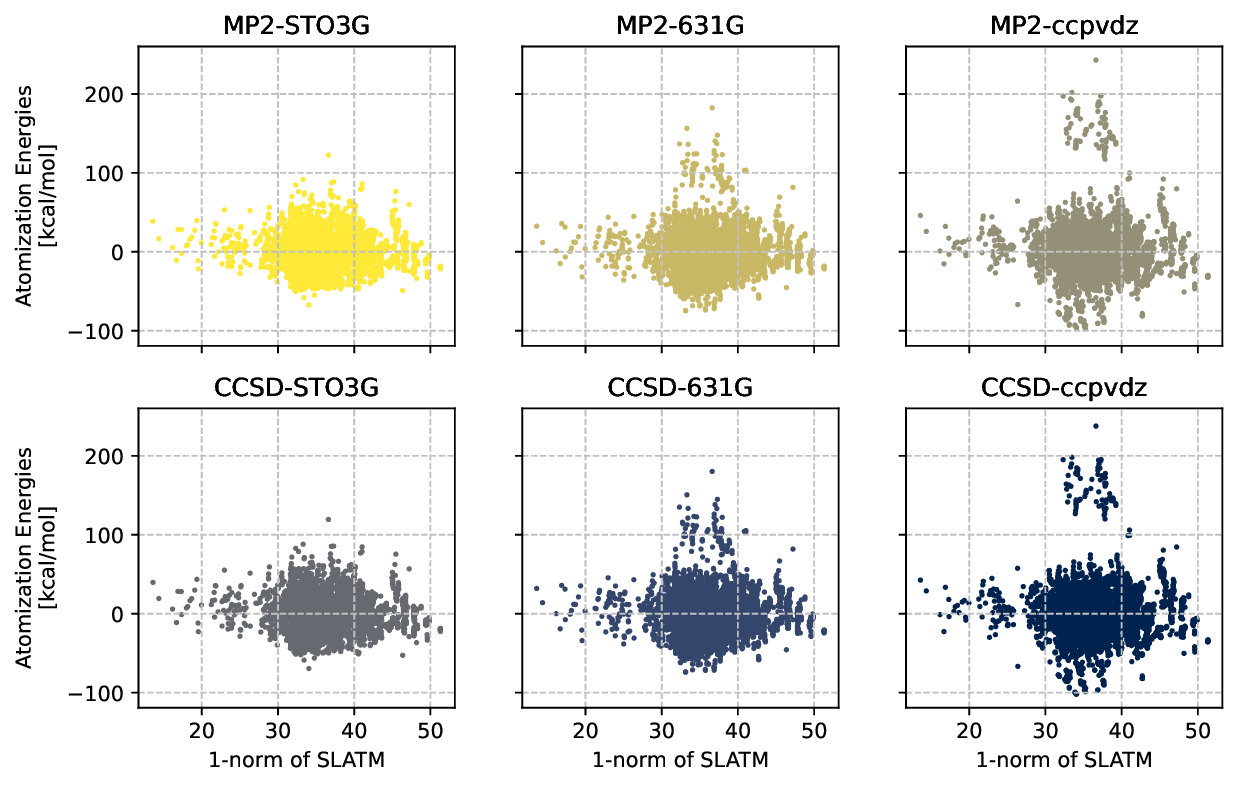}
    \caption{Scatter plot of the various fidelities from the training data with respect to the 1-norm of the corresponding SLATM representation. The SLATM representation serves as a proxy to the chemical-space. Thus these scatter plots represent the spread of the atomization energies across the chemical space. The first row corresponds to the MP2 level of theory for increasing basis set sizes. Similarly, the second row displays the scatter plots for the CCSD level of theory.}
    \label{fig_SLATM_Scatter}
\end{figure}
Previous work by Zaspel \textit{et.~al.}~already provided a benchmark for the MFML method in prediction of the atomization energies for various molecules in the QM7b dataset \cite{zasp19a}. The same parameters of the Laplacian kernel of width 400, and regularization of $10^{-10}$ are used in this work to maintain uniformity for comparison of MFML and o-MFML.
The updated o-MFML models are benchmarked on the same dataset with modifications as reported in Section \ref{Dataset}. The hyper-parameters of KRR are chosen to be identical to the values reported in the previous work. In total, 6 fidelities are considered with the target fidelity of CCSD-ccpvdz being the costliest and the MP2-STO3G being the cheapest fidelity. Further details are discussed in Section \ref{Dataset}.

As a preliminary analysis, the scatter plot between the 1-norm of SLATM representations \cite{Huang2020slatm} and the atomization energies of the molecules from the training set is studied in Figure \ref{fig_SLATM_Scatter}.
This assists in understanding the layout of the chemical space by studying the proxy of the chemical space, which in this case is the SLATM representation. On comparing the distribution across the basis sets, that is, row-wise, one observes that increasing basis set size results in clearer separation of the atomization energies across the proxy chemical space. The higher energy clusters become clearer.
A similar comparison for increasing level of theory shows visible differences only for the ccpvdz basis set. 
Here, the CCSD level of theory further separates clusters of molecules in comparison to the MP2 level of theory, especially for those with atomization energies in the region of -100 kcal/mol. 
For increasing accuracy to the target fidelity of CCSD-ccpvdz, one observes that the scatter plot of the energies with respect to the chemical space gets closer to that of the target fidelity. 
The smallest basis set, STO3G does not show any atomization energies higher than 100 kcal/mol for both MP2 and CCSD levels of theory. One observes that each increasing fidelity results in a clearer, more distinct categorization of the molecules in the QM7b dataset, which was previously discussed in Ref.~\cite{zasp19a} with respect to the 1-norm of the coulomb matrices. 
The STO3G basis sets fail to provide any form of information of the separation of the clusters of molecules. The scatter plot of the fidelities with this basis set show a strong clustering around the 0 kcal/mol mark. For the larger basis sets, one observes that higher atomization energies show two distinct clusters. A large one around the 0 kcal/mol mark and another around the 150 kcal/mol mark. As identified in Ref.~\cite{zasp19a}, these correspond to the largest molecules of the QM7b dataset. 
Since this information is missing from the smaller STO3G basis set, one anticipates that the use of the fidelities MP2-STO3G and CCSD-STO3G in the conventional MFML would provide little to no benefit in predicting the atomization energies at the target fidelity of CCSD-ccpvdz where the clustering is all the more distinct.

\begin{figure}
    \centering
    \includegraphics[width=0.75\textwidth]{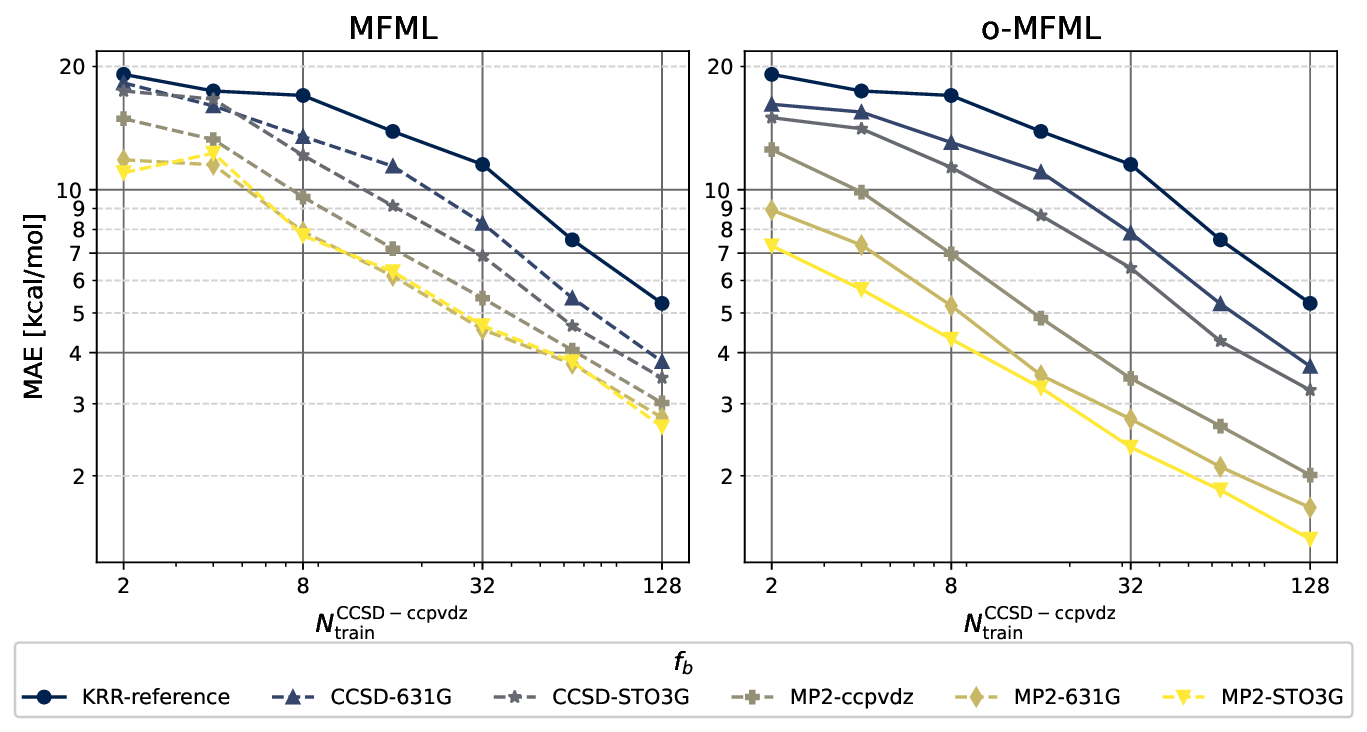}
    \caption{Various learning curves for the prediction of atomization energies of molecules in the QM7b dataset. The left-hand side plot corresponds to learning curves built with the conventional MFML method, that is $P_{\rm MFML}^{\rm (F;f_b)}$. 
    The right-hand side plot corresponds to the o-MFML models optimized with OLS, referred to as $P_{o-MFML}^{\rm (F;f_b)}$.
    In both cases, each curve corresponds to a model where the target fidelity, $F$, is CCSD-ccpvdz. The various baseline fidelities $f_b$ are as shown in the figure legend. The learning curve for the conventional KRR model (KRR-reference) is also shown for reference.
    }
    \label{fig_SLATM_LC}
\end{figure}
The resulting learning curves of the multifidelity analysis on the QM7b data are shown in Figure \ref{fig_SLATM_LC}. All the sub-models for MFML and o-MFML methods were built with KRR using the Laplacian Kernel with regularization strength of $10^{-10}$ and a kernel width of 400 as prescribed in Ref.~\cite{zasp19a}.
The left hand side of the figure depicts the learning curves for the conventional MFML method with default coefficients for the sub-models. The learning curves for o-MFML  method are depicted in the right-hand side of the same figure. The conventional reference kernel ridge regression (KRR) learning curve is presented in both panes for ready reference. The horizontal axis denotes the number of training samples used at the target fidelity in training the various models. 
For conventional MFML learning curves, one observes distinct lowered offsets of the learning curves with decreasing baseline fidelities. As preemptively discussed in the preliminary analysis, the addition of MP2-STO3G fidelity does not provide any perceivable benefit to the MFML model. The model built on the CCSD-STO3G baseline, however, does show improvement. 

The learning curves for the o-MFML models are presented on the right-hand side of Figure \ref{fig_SLATM_LC}. Firstly, one observes that even for smaller training set sizes, the o-MFML does not show any pre-asymptotic perturbance. The MAE of the various models always decreases for increasing training samples. This is contrasted with the conventional MFML method where a region of pre-asymptotics is observed wherein the MAE of the model built with $f_b=$~MP2-STO3G fluctuates before settling down.
In other words, there is a constantly lowered offset with the addition of each cheaper fidelity even for very small training set sizes for the o-MFML models.  
The same sub-models are used for both MFML and o-MFML models. The combination of these models is optimized resulting in an increased accuracy of prediction.
Secondly, one also notices that the addition of the MP2 level of theory even with the largest basis set size results in a significant decrease in the prediction error of the model. The addition of the MP2-STO3G fidelity further improves the capability of predictions of the o-MFML models resulting in a lower error of prediction. For $N_{\rm train}^{\rm CCSD-ccpvdz}=128$ and the baseline of MP2-STO3G, the MFML method results in an MAE of 2.73 kcal/mol while the MAE corresponding to the o-MFML method is 1.4 kcal/mol. 

\begin{figure}
    \centering
    \includegraphics[width=0.75\textwidth]{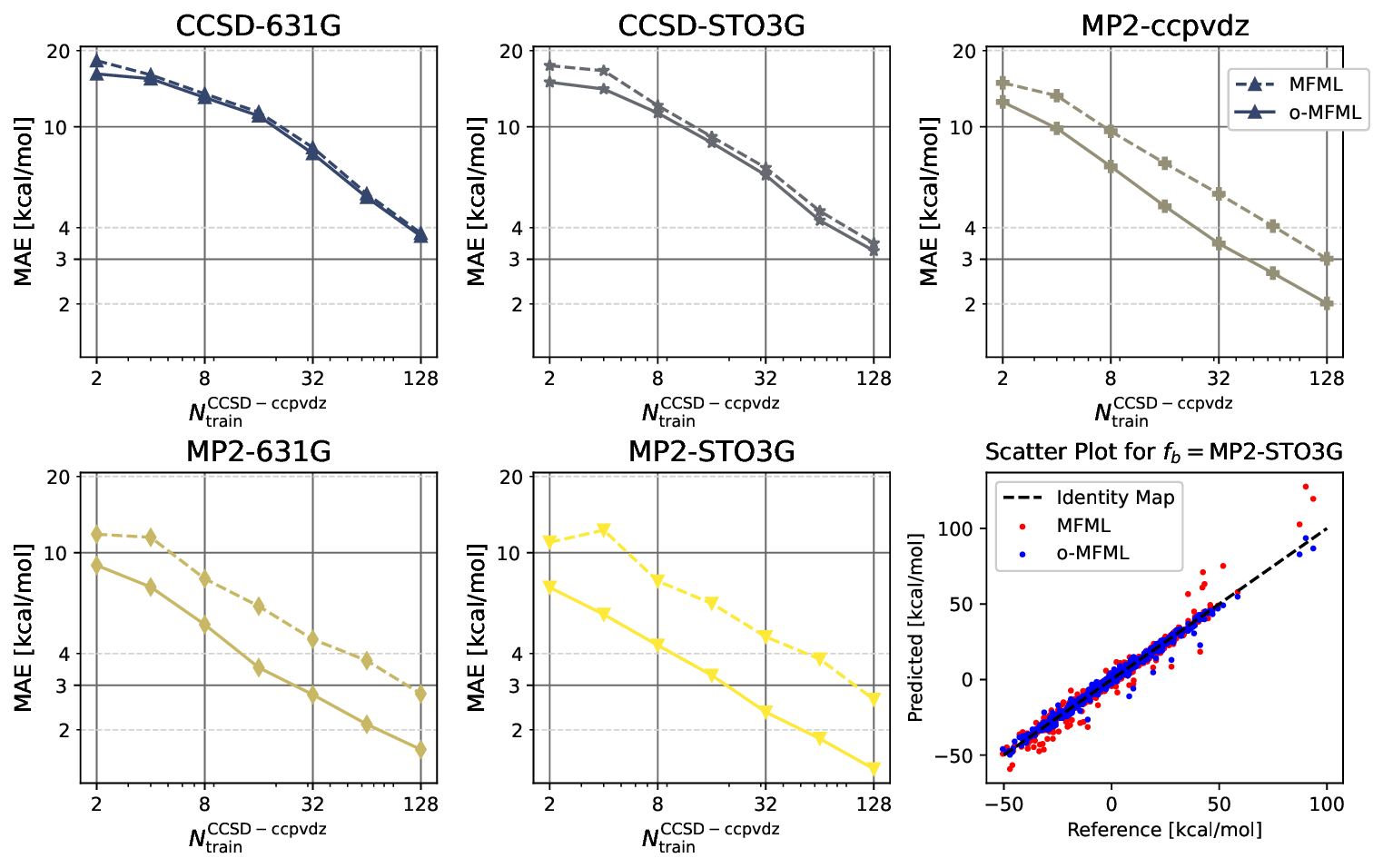}
    \caption{
    A comparison of learning curves of the MFML (dashed lines) and o-MFML (solid lines) models for varying baseline fidelities, $f_b$. A scatter plot comparing the predictions vs. CCSD-ccpvdz reference for the two models is also presented for $f_b=$ MP2-STO3G.
    }
    \label{fig_singleview_scatter}
\end{figure}

The improvements offered by the o-MFML method become more evident when one compares individually the learning curves of the MFML and o-MFML models for each baseline fidelity. This is done in Figure \ref{fig_singleview_scatter}, where, for each baseline fidelity, $f_b$, the learning curves of the MFML and o-MFML are compared for decreasing fidelity. Already for the baselines from the CCSD level of the theory the improvement of the o-MFML method is visible, but not significantly. The stark decrease of the MAE with the addition of MP2-ccpvdz fidelity becomes evident in the right-most pane in the first row of this figure. Subsequent addition of cheaper fidelities further reduces this offset in comparison to the conventional MFML method. For the case of the MP2-STO3G baseline, the o-MFML method for the predictions of atomization energies results in an MAE that is almost twice as lower when compared to the MFML method. It becomes evident that for each case of the baseline, the o-MFML models are superior predictors in comparison to the conventional MFML methods. 

A closer look at this model for $N_{\rm train}^{\rm CCSD-ccpvdz}=2^7=128$ clarifies this interpretation. The last pane in the second row of Figure \ref{fig_singleview_scatter} shows the scatter plot between the reference atomization energies of the molecules from the test set and the prediction of the two multifidelity methods on the same molecules. Identical training data was used to build the various sub-models for both the MFML and o-MFML methods. 
One immediately observes that the spread for the o-MFML model in the scatter plot is closer to the identity mapping than that for the MFML model. Of particular interest are the areas around -50 kcal/mol and beyond 50 kcal/mol. The MFML model consistently under-estimates the atomization energies at the lower end while over-estimating those at the upper end of the energy range. The over-estimation in particular begins as early as about 40 kcal/mol and becomes evident as one goes up the energy range. The o-MFML on the other hand manages to predict these higher atomization energies with higher accuracy thus bringing the distribution closer to the identity mapping. 

\subsubsection{Coefficient Study}\label{coeff_analysis_res}
\begin{figure}
    \centering
    \includegraphics[width=0.75\textwidth]{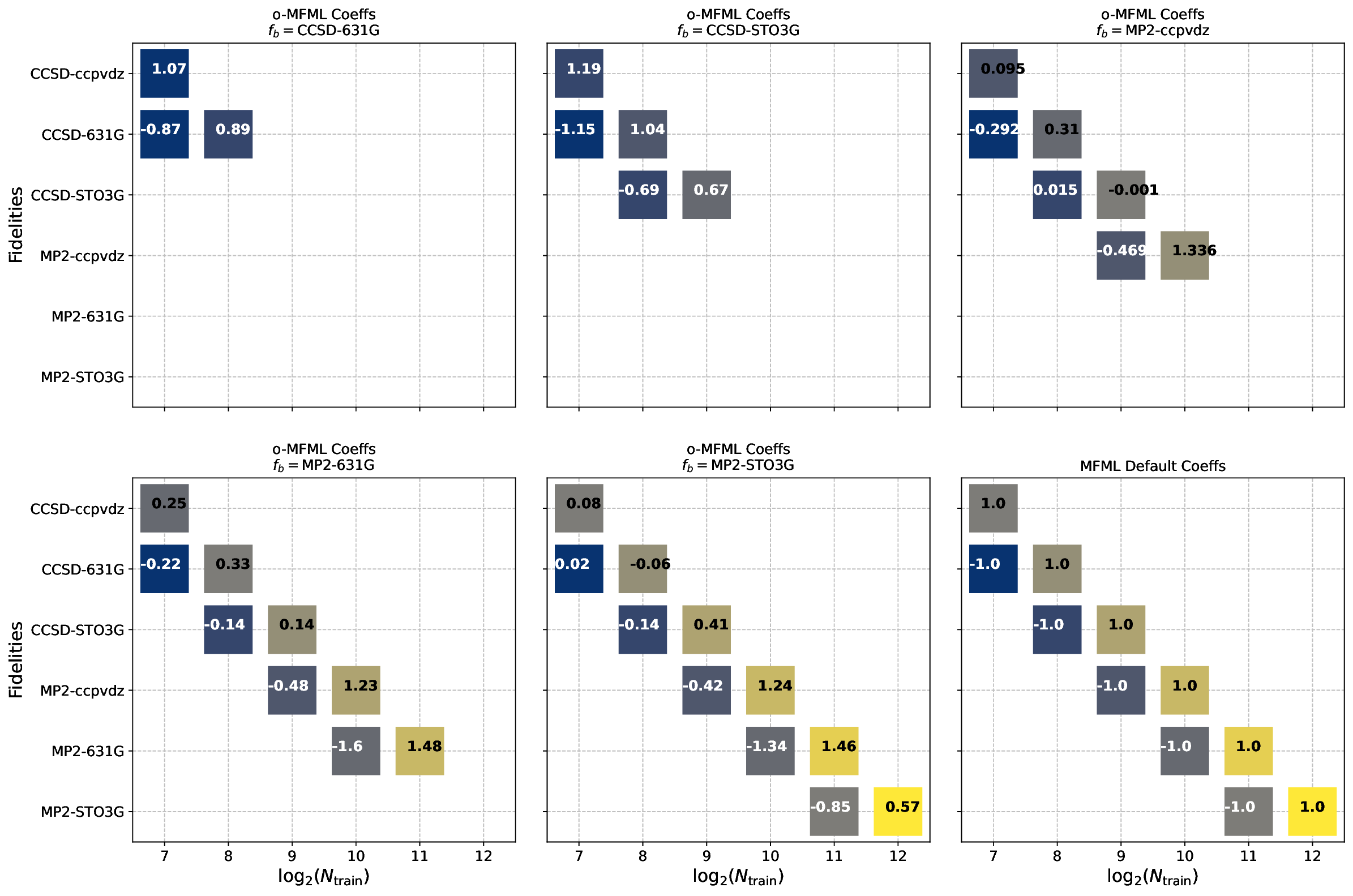}
    \caption{
    Values of the o-MFML coefficients for $N_{\rm train}^{\rm CCSD-ccpvdz}=2^7=128$.
    For readability, in most of the cases, the coefficients have been rounded off to the second decimal place. For varying baseline fidelities, the final values of the coefficients are shown. For reference, the default coefficients used in MFML are shown for the MP2-STO-3G baseline.
    }
    \label{fig_SLATM_coeffs}
\end{figure}

As discussed in Section \ref{POM_details}, the o-MFML method optimally combines the various sub-models to result in a superior multifidelity method. The coefficients are optimized on the validation set with the OLS method. In order to further understand the o-MFML method, the analysis of these coefficients is performed as seen in Figure \ref{fig_SLATM_coeffs}. 
The default coefficients used in the MFML methods are depicted in the last column of the second row. Notice that this corresponds to the discussion in Section \ref{MFML_methods} wherein the MFML model is built with the differences between the sub-models. 
For the different o-MFML models, one observes that the coefficients of each sub-model, $P_{\rm KRR}^{\boldsymbol{s}}$, vary with varying baseline fidelities. This change signifies the optimization of the MFML model with respect to the validation set. 

A meaningful analysis of the different cases is the comparison of the magnitude of the coefficients $\beta_{\boldsymbol{s}}^{\rm opt}$ to $\beta_{\boldsymbol{s}}^{\rm MFML}$. For the o-MFML models built for baseline fidelities from the CCSD level of theory, the coefficients are close in magnitude to those of the conventional MFML. This could imply that the MFML method was already nearly optimized for these fidelities.
With the addition of the MP2 fidelities, however, the coefficient landscape changes. The optimization of the coefficients results in values that are significantly different from the conventional $\beta_{\boldsymbol{s}}^{\rm MFML}$ values. This flexibility in combining sub-models rather than simply adding the differences (as done in MFML) allows o-MFML to be a superior method. 
The middle and right-hand side plots of the second row in Figure \ref{fig_SLATM_coeffs} assist in comparing the values of $\beta_{\boldsymbol{s}}^{\rm MFML}$ and $\beta_{\boldsymbol{s}}^{\rm opt}$ for the case of the MP2-STO3G baseline. There is significant difference in the optimized coefficients and the default MFML coefficients for almost all the sub-models. This shows that the conventional MFML method was not optimized in combining the different fidelities. 

In particular one observes that the values of $\beta_{\boldsymbol{s}}^{\rm opt}$ for the CCSD-631G fidelity are small in comparison with those of the other fidelities in the central plot of the second row. This could indicate that the optimization method identified this fidelity to be less useful.
In order to verify this, an experiment was carried out by separately building two models. The first was the usual complete model with all six fidelities with $N_{\rm train}^{\rm CCSD-ccpvdz} = 2^7$ and the training samples at the other fidelities scaled by 2. The second model was built without the CCSD-631G fidelity but the training samples at the other fidelities were kept to be identical to that used in the first model, that is, $(2^7,2^{9},2^{10},2^{11},2^{12})$. For these two models the o-MFML was generated and the MAE evaluated. The original model resulted in an MAE of 1.421 kcal/mol while the second model resulted in an MAE of 1.431 kcal/mol which is a difference of 0.72\%. 
This is a strong indicator towards the robustness of the o-MFML method and how it can be a tool to detect whether a particular fidelity would benefit the overall multifidelity structure or not. More details on the effectiveness of the coefficient analysis are reported in the supplementary material in Section 3.3.1.

\subsection{Excitation  Energy Prediction}\label{excitedstate_results}
\begin{figure}
    \centering
    \includegraphics[width=0.75\textwidth]{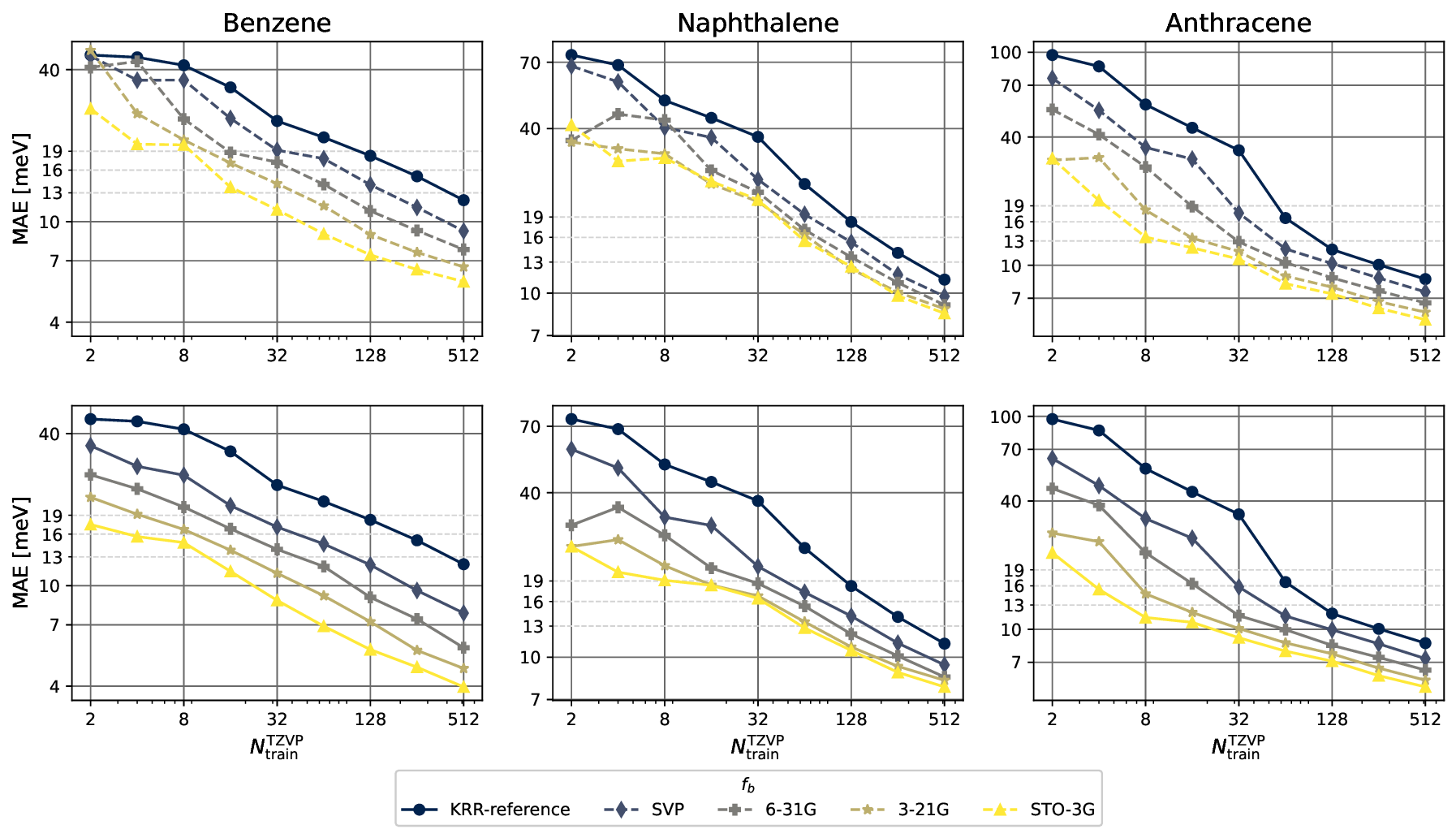}
    \caption{Learning curves for MFML (top row) and o-MFML (bottom row) models for MD-based trajectories of various molecules for the prediction of excitation  energies. The various baselines fidelities used are delineated in the legend. The KRR-reference (black curve) is provided for each case for a single-fidelity training on TZVP. The axes are scaled identically for the MFML and o-MFML methods but are different for each of the different molecules.}
    \label{fig_excited_CMD_LC}
\end{figure}

The dataset for excitation  energies consists of MD and DFTB-based trajectories of benzene, naphthalene, and anthracene \cite{vinod23_MFML}. A total of 5 fidelities were calculated and ordered as discussed in Section \ref{Dataset}. In brief, the target fidelity is set to be TZVP and the cheapest fidelity is considered to be STO-3G. All the sub-models used in both the MFML and o-MFML method are built with KRR using the Mat\'ern Kernel of first order and $l_2$ norm. A regularization strength of $10^{-9}$ is used. The kernel widths for each molecule were chosen as recorded in Ref.~\cite{vinod23_MFML}. Unsorted coulomb matrices are used as representations for all cases.
Previously, various preliminary analyses of this dataset have been discussed and two problematic data structures were thereby identified \cite{vinod23_MFML}. 
For MD-based naphthalene, there was no clear multifidelity structure. For DFTB-based anthracene, a high spread of the STO-3G energies with respect to the target fidelity of TZVP was also identified to be problematic. From these, it was shown that the MFML method would not provide favorable results for these two cases. 

The learning curves of the conventional MFML method for the MD-based trajectories of benzene, naphthalene, and anthracene are shown in the top row of Figure \ref{fig_excited_CMD_LC}. At the same time, the bottom row shows the learning curves resulting from the novel o-MFML method. Various baselines fidelities for the multifidelity models are as shown in the legend. 
Of particular interest in this is the case of naphthalene. The MFML results reflect the issue of the wide spread of the scatter as previously identified in Ref.~\cite{vinod23_MFML}. However, with the o-MFML method, one observes that the model built with the 3-21G and 6-31G fidelities still results in constant lowered offsets as opposed to the conventional MFML method where these models do not provide much improvement. Thus, the o-MFML method provides a robust multifidelity method even if the data distribution of the quantum chemistry methods is not as anticipated for MFML. For benzene and anthracene, the improvement in the MAEs is perceptibly small. This could indicate that the original MFML model already was properly optimized for these cases. 

\begin{figure}
    \centering
    \includegraphics[width=0.75\textwidth]{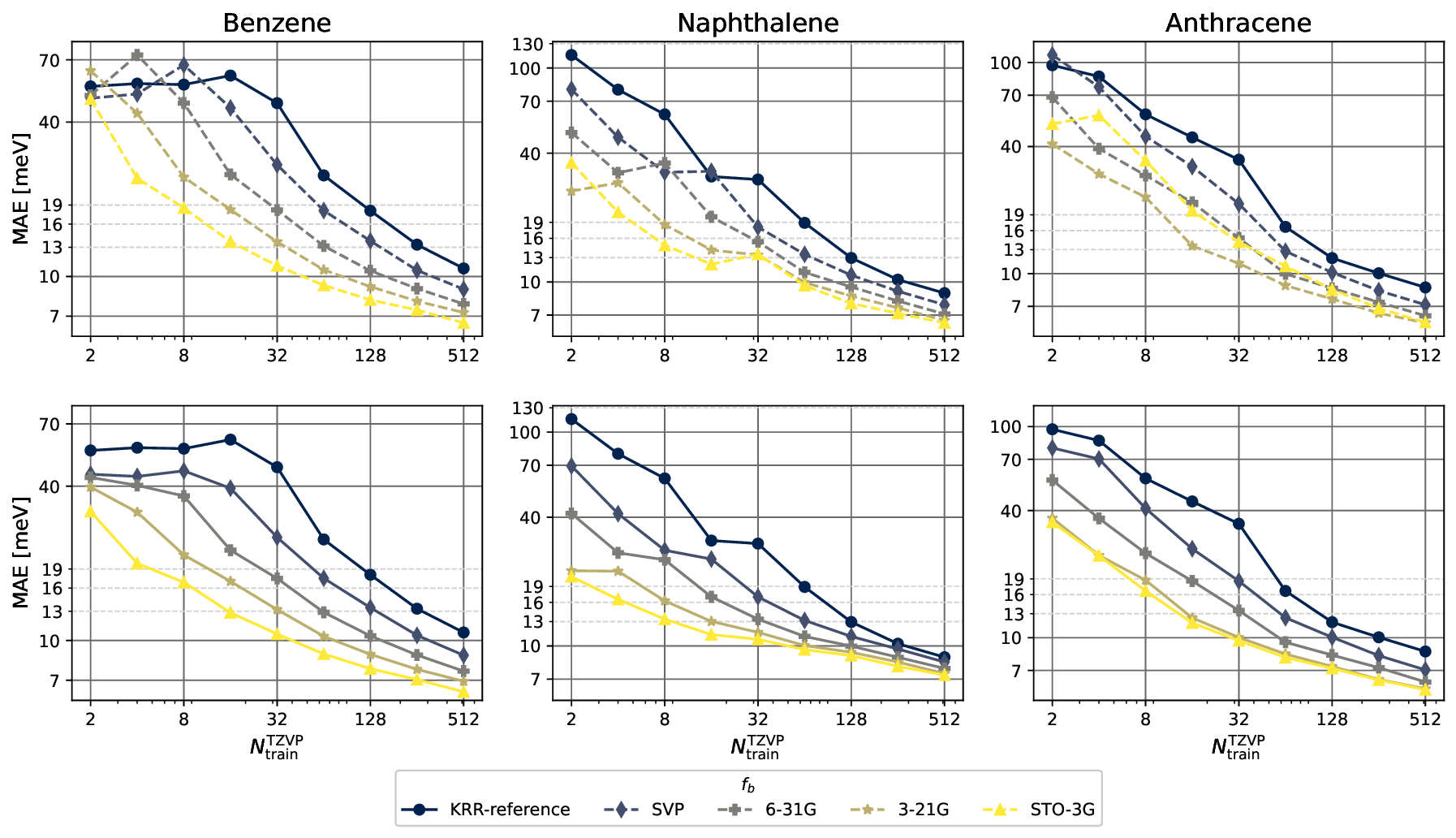}
    \caption{Learning curves for MFML (top row) and o-MFML (bottom row) for DFTB-based trajectories of various molecules. The MAE is reported for the prediction of first excitation  energies. The single-fidelity (TZVP) KRR leaning curve (black line) for prediction on the same test set as the other models is provided for reference. The scaling of the axes is identical for individual molecules across the MFML and o-MFML models for easy comparison.}
    \label{fig_excited_DFTB_LC}
\end{figure}
Similarly, the learning curves for the DFTB-based trajectories of various molecules are given in Figure \ref{fig_excited_DFTB_LC}. 
As for the case of the MD trajectories, the use of the o-MFML method results in models which perform consistently better across the various training set sizes. That is, even for smaller training set sizes, the benefit of the multifidelity structure becomes evident. 
Across the molecules, for smaller training set sizes, the learning curves for the MFML methods have various cross overs which indicates a distinct region of pre-asymptotics for smaller training set sizes. That is, the benefit of using cheaper baselines does not become evident until sufficiently large training samples are used.
In contrast, the o-MFML method appears to have smoothed out any pre-asymptotics. even for very small training set sizes, the addition of each cheaper baseline shows immediate decrease in the MAEs of the models. 
Next, consider the case of DFTB-based anthracene. For the MFML method, the addition of the STO-3G fidelity results in a decrease in performance of the model as discussed in Ref.~\cite{vinod23_MFML}, where the authors argue that the wide spread distribution of the STO-3g fidelity with respect to the target fidelity of TZVP results in a poorer improvement with the conventional MFML method.
With the o-MFML method, the optimization of the coefficients results in a model that performs much better. The learning curve indicates that the o-MFML model for the STO-3G baseline is now comparable to that of the model built with the 3-21G baseline. 
The o-MFML method results in a better model in-spite of the poor distribution of the STO-3G with respect to the target fidelity. Further results and analyses of the o-MFML employed for the prediction of the excitation  energies are discussed in S~3.3.

\section{Conclusion}
This work has numerically established the improvement of the conventional MFML by optimally combining the various multifidelity sub-models. 
For the prediction of atomization energies of molecules from the QM7b dataset, and the prediction of excitation  energies for three molecules of growing sizes, o-MFML has been shown to categorically improve the prediction capabilities of the multifidelity method. 
The use of o-MFML was especially shown to be beneficial for cases where the hierarchy or distribution of the cheaper fidelities is not optimal. 
The learning curves indicate that the use of o-MFML results in low errors for the prediction of both atomization energies and excitation  energies. This novel method opens up further research avenues for multifidelity methods in QC. The use of the optimal coefficients to determine the optimal number of training samples to be used at each fidelity, for instance, is one such area of research. When combined with an in-depth analysis of the scaling of the number of training samples between fidelities, this could provide a better picture of the multifidelity structure and its use for QC properties. Overall, this work presents a cost-efficient and optimized multifidelity model with promising outlook in QC applications.

\section*{Associated Content}
Supplementary Material sections S1-S3, Figs.~S1-S4, and Table S1. 

\section*{Acknowledgments}
The authors acknowledge support under the DFG grant 466761712 for the project ``Excitation Energy Transfer in a Photosynthetic System with more than 100 Million Atoms".


\section*{Conflict of Interest}
The authors declare no conflict of interest.

\section*{Ethical statement}
This work does not include human subjects, human data or tissue, or animals.

\bibliography{bibvivinod}

\end{document}